\documentclass[a4paper]{jpconf}
\usepackage{graphicx}

\newcommand{\beq}{\begin{equation}}
\newcommand{\eeq}{\end{equation}}
\newcommand{\beqar}{\begin{eqnarray}}
\newcommand{\eeqar}{\end{eqnarray}}
\newcommand{\ds}{\displaystyle}
\newcommand{\bec}{\begin{center}}
\newcommand{\enc}{\end{center}}

\begin{document}
\title{HYDRO + JETS (HYDJET++) event generator for Pb+Pb collisions at
       LHC}

\author{L~Bravina$^{1,4,5}$, B~H~Brusheim~Johansson$^1$, 
J~Crkovsk\'{a}$^3$, G~Eyyubova$^2$, V~Korotkikh$^2$, I~Lokhtin$^2$, 
L~Malinina$^2$, E~Nazarova$^2$, S~Petrushanko$^2$, A~Snigirev$^2$, 
E~Zabrodin$^{1,2,4,5}$
}

\address{$^1$
Department of Physics, University of Oslo, Norway}
\address{$^2$
Skobeltsyn Institute of Nuclear Physics, Lomonosov Moscow State 
University, Moscow, Russia}
\address{$^3$
Institut de Physique Nucl\'{e}aire, CNRS-IN2P3,
Univ. Paris-Sud, Universit\'{e} Paris-Saclay, France}
\address{$^4$
Frankfurt Institute for Advanced Studies, D-60438 Frankfurt a.M.,
Germany}
\address{$^5$
National Research Nuclear University "MEPhI"
(Moscow Engineering Physics Institute), Moscow, Russia}

\ead{larissa.bravina@fys.uio.no}

\begin{abstract}
The Monte Carlo event generator HYDJET++ is one of the few generators,
designed for the calculations of heavy-ion collisions at 
ultrarelativistic energies, which combine treatment of soft hydro-like
processes with the description of jets traversing the hot and dense 
partonic medium. The model is employed to study the azimuthal 
anisotropy phenomena, dihadron angular correlations and event-by-event
(EbyE) fluctuations of the anisotropic flow in Pb+Pb collisions
at $\sqrt{s_{\rm NN}} = 2.76$~TeV. 
The interplay of soft and hard processes describes the violation of the 
mass hierarchy of meson and baryon elliptic and triangular flows at
$p_{\rm T} \geq 2$~GeV/$c$, the fall-off of the flow harmonics at 
intermediate transverse momenta, and the worsening of the 
number-of-constituent-quark (NCQ) scaling of elliptic/triangular flow 
at LHC compared to RHIC energies. The cross-talk of $v_2$ and $v_3$ 
leads to emergence of higher order harmonics in the model and to 
appearance of {the \it ridge} structure in dihadron angular correlations 
in a broad pseudorapidity range. HYDJET++ possesses also the dynamical 
EbyE fluctuations of the anisotropic flow. The model results agree well 
with the experimental data. 
\end{abstract}

\section{Introduction. HYDJET++ model}
\label{intro}

The Monte Carlo event generator HYDJET++ (HYDrodynamics with JETs)
\cite{hydjet++} consists of two parts describing the soft and the hard
processes, respectively. The generator of soft processes FASTMC 
\cite{fastmc1} was originally designed to help the experimentalists 
in simulation of significant data samples of heavy-ion collisions at
energies of RHIC and LHC. The demands to the model were formulated
as follows: (i) it should be able to simulate tens of thousands of 
central gold-gold or lead-lead events at energies from $\sqrt{s_{\rm NN}} 
= 200$~GeV to several TeV within relatively modest CPU time; (ii) the 
yields and transverse momentum spectra of most abundant particles 
should be very close to real data. To meet these requirements the
authors opted for parameterised ideal hydrodynamics. The model was 
extended soon to non-central nuclear interactions \cite{fastmc2}.
The first step is the calculation of the effective volume of the 
fireball $V_{\rm eff}$ which is a subject of the mean number of
participating nucleons at given collision centrality. Particle 
composition is frozen at chemical freeze-out, but the fireball expands
further and breaks down at thermal freeze-out temperature, where the
contact between hadrons is lost. The final-state interactions (FSI)
assume two- and three-body decays of resonances. The table of particles 
in the model contains more than 360 meson and baryon states including 
also the charmed ones. 
This approach is close to the THERMINATOR model \cite{therm}. Obviously,
the ideal hydrodynamic description of particle spectra is justified for 
transverse momenta below 2~GeV/$c$. At higher $p_{\rm T}$ one 
has to take into account hard processes. These processes are governed
by the PYQUEN routine \cite{pyquen}, which propagates the hard partons
through the hot and dense medium, most presumably quark-gluon plasma.
The partons experience collisional and radiative energy losses. At the
end of the rescattering stage all emitted quarks and gluons are
hadronized according to the Lund model. The number of produced jets in 
HYDJET++ is proportional to the number of binary nucleon-nucleon
collisions at a given impact parameter and the integral cross section 
of the hard processes in NN collision with minimal transverse momentum 
transfer.

The synergy between the soft processes and quenched jets became obvious 
soon after the merging of two independent generators, FASTMC and PYQUEN, 
into the model called HYDJET++ \cite{hydjet++}. Several examples 
concerning the interplay of hard and soft processes will be discussed in 
Sec.~\ref{sec2}. Then, the model was further upgraded. Namely, after the 
publication of first data on proton-proton collisions at 
$\sqrt{s_{\rm NN}} = 7$~TeV it became clear that the standard version of 
PYTHIA\_6.4 \cite{pythia6.4} should be adjusted. Several tunes have been 
proposed and the HYDJET++ group has opted for Pro-Q20 tune in the new 
release of the model \cite{hj_12}. Next was the extension of the model 
to triangular flow \cite{hj_hiharm}.

Recall that the azimuthal distribution of particles can be cast
\cite{VoZh96,PoVo98} in the form of Fourier series
\beq
\ds
E \frac{d^3 N}{d^3 p} = \frac{1}{2 \pi} \frac{d^2 N}{p_{\rm T}dp_{\rm T} 
dy} \left\{ 1 + \sum_{n=1}^{\infty} 2 v_n \cos{\left[n(\phi - \Psi_n)
\right] } \right\}~.
\label{eq1}
\eeq
Here $\phi$, $p_{\rm T}$ and $y$ are the azimuthal angle, the transverse 
momentum and the rapidity of a particle, respectively. $\Psi_n$ is the
azimuth of the corresponding event plane, and the sum
of harmonics in the rhs of Eq.~(\ref{eq1}) represents anisotropic flow. 
The coefficients $v_n$ are dubbed directed flow $v_1$, elliptic flow 
$v_2$, triangular flow $v_3$ and so forth. In HYDJET++ elliptic and 
triangular flows arise because of the corresponding spatial 
eccentricities of the fireball.
The radii of the elliptic and triangular spatial eccentricities are 
defined as function of the impact parameter $b$ and azimuthal angle 
$\phi$ as follows \cite{fastmc2,hj_hiharm} :
\beqar
\ds
R_{\rm ell}(b, \phi) &=& R_{\rm fr} \left[ \frac{1 - \varepsilon^2(b)}
{1 + \varepsilon^2(b) \cos{2(\phi - \Psi_2)}} \right]^{1/2}~,\\
\label{eq2}
R_{\rm trian}(b,\phi) &=& R_{\rm ell}(b, \phi) \{ 1 + \varepsilon_3(b) 
\cos{3(\phi - \Psi_3)} \}~,
\label{eq3}
\eeqar
where $R_{\rm fr}(b) = R_0 \sqrt{1 - \varepsilon(b)}$, and $R_0$ is the
freeze-out radius of the fireball in a central collision. Two free 
parameters, $\varepsilon(b)$ and $\varepsilon_3(b)$, control the
ellipticity and triangularity of the fireball. $\Psi_2$ and $\Psi_3$ are 
the azimuths of the corresponding event planes randomly distributed with
respect to each other. The pressure gradients 
are stronger in the direction of short axis of the ellipsoid, however,
the momentum anisotropy angle $\phi_{\rm fl}$ is related to the spatial
anisotropy $\phi$ via
\beq
\ds
\frac{\tan{ \phi_{\rm fl}}}{\tan{\phi}} = \left[ \frac{1 - \delta(b)}
{1 + \delta(b)} \right]^{1/2}~,
\label{eq4}
\eeq     
containing the third (and the last) free parameter, $\delta(b)$, 
responsible for the formation of anisotropic flow in HYDJET++. 
The cross-talk of elliptic and triangular harmonics leads to appearance 
of both even and odd higher harmonics of the anisotropic flow in the 
model because of the nonlinear contributions to $v_n$ from $v_2,\ v_3$
or their product $v_2 v_3$. This interplay explains also the formation
of ridge in long-range dihadron correlations, including the 
characteristic double-bump profile of the ridge at the away-side
\cite{ridge_prc15}. The last developments of the model deal with its 
extension to open and hidden charm production \cite{charm_arx} and to 
EbyE fluctuations of the anisotropic flow \cite{fluct_epjc15}. The 
aspects of the calculations are presented in Sec.~\ref{sec2}. 
Conclusions are drawn in Sec.~\ref{concl}.  

\section{Interplay of soft processes, jets and final-state interactions}
\label{sec2}

{\it Consequences for elliptic and triangular flows.}
Hadrons produced in soft processes at the freeze-out hypersurface should 
carry collective flow, whereas the flow assigned to jet particles with 
intermediate transverse momenta is essentially zero. At 
$p_{\rm T} \geq 4$~GeV/$c$ hadrons decoupled from jets can develop a 
weak anisotropic flow because of the well-known effect of jet quenching. 
Decays of resonances also modify the $p_{\rm T}$ distributions of the 
flow excitation functions. In HYDJET++ these effects were studied in 
non-central heavy-ion collisions at RHIC and LHC energies in 
\cite{v2_prc09,v2_sqm09} for the elliptic flow and in
\cite{v3_sqm15,v3_arxiv} for the triangular flow, respectively.

\begin{figure}[h]
\bec
\vspace{0.5cm}
\includegraphics[scale=0.60]{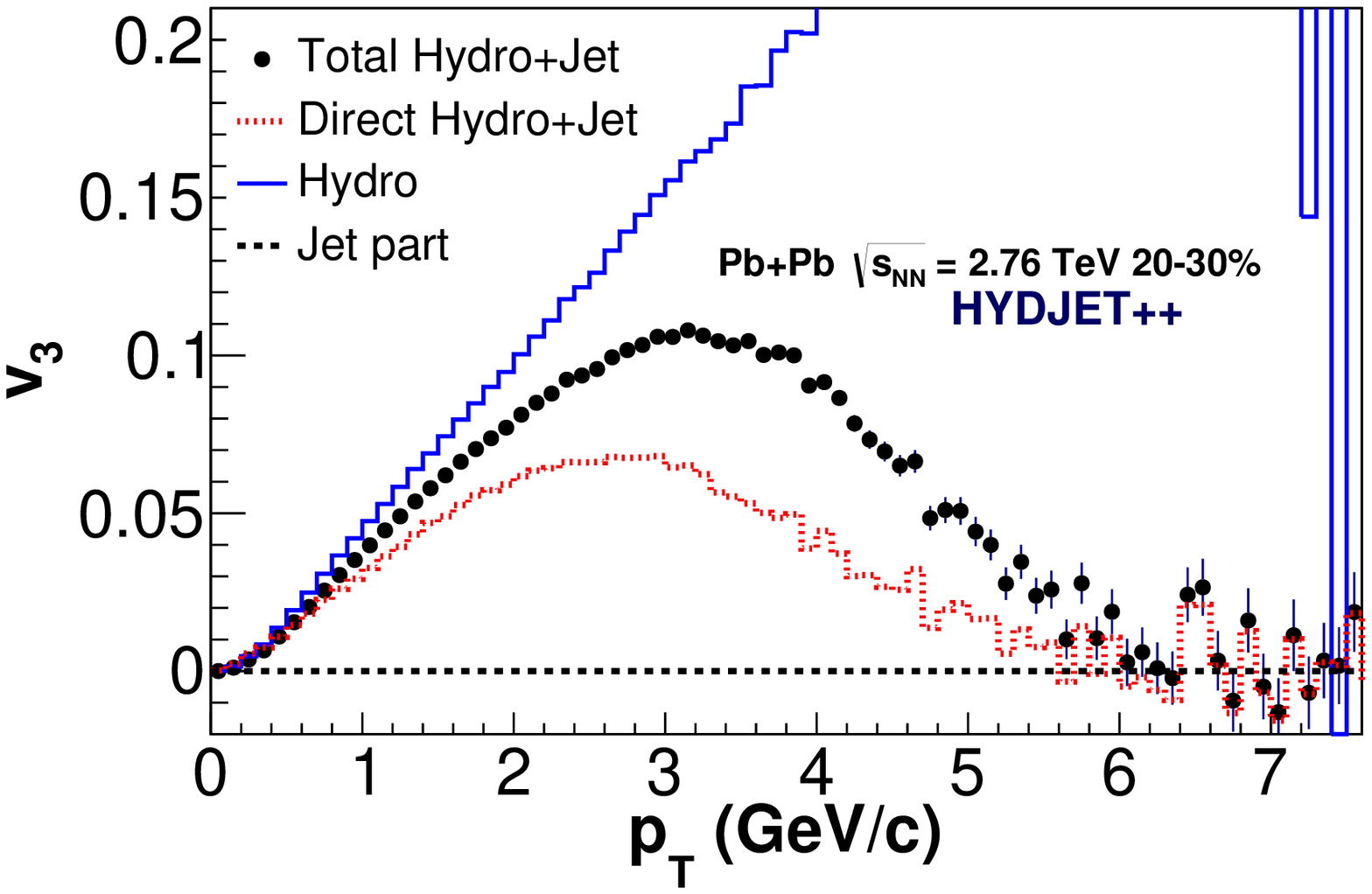}
\caption{\label{fig1}
The $p_{\rm T}$ dependence of the triangular flow of charged particles 
produced in soft+hard (circles) and soft processes (solid line), from 
the jets (dashed line) and direct particles (dotted line) in HYDJET++ 
in Pb + Pb collisions at $\sqrt{s_{\rm NN}} = 2.76$~TeV at centrality
20-30\%.}
\enc
\end{figure}

To illustrate the typical features of the development of the flow 
components we plot in Fig.~\ref{fig1} triangular flow of charged hadrons
produced in Pb+Pb collisions at $\sqrt{s_{\rm NN}} = 2.76$~TeV at
centrality $\sigma /\sigma_{geo} = 20 - 30\%$. The partial flow of jet 
hadrons is absent, while the flow of particles governed by hydrodynamics 
increases with rising $p_{\rm T}$. Hadrons from the soft processes 
dominate the particle spectrum at $p_{\rm T} \leq 2$~GeV/$c$, and the 
total $v_3$ increases in this $p_{\rm T}$ range. At higher transverse 
momenta the fraction of jet hadrons prevails over the ``soft" ones. 
Thus, the triangular flow in HYDJET++ decreases after a certain
$p_{\rm T}$. The lighter the hadron, the smaller is the value of 
transverse momentum where the spectra of hadrons originated from the 
soft and the hard processes cross each other. This circumstance explains 
the violation of the mass ordering of hadron elliptic and triangular
flows. One can also see in Fig.~\ref{fig1} that the decays of resonances 
increase the maximum of the $v_3(p_{\rm T})$ distribution by about 20\% 
and shift its position to higher $p_{\rm T}$. However, 90\% of hadrons
have the transverse momenta less than 1~GeV/$c$; therefore, the
$p_{\rm T}$-integrated values of $v_3$ at selected centralities are 
changed insignificantly \cite{v3_arxiv}.

\begin{figure}[h]
\bec
%\vspace{0.5cm}
\includegraphics[scale=0.50]{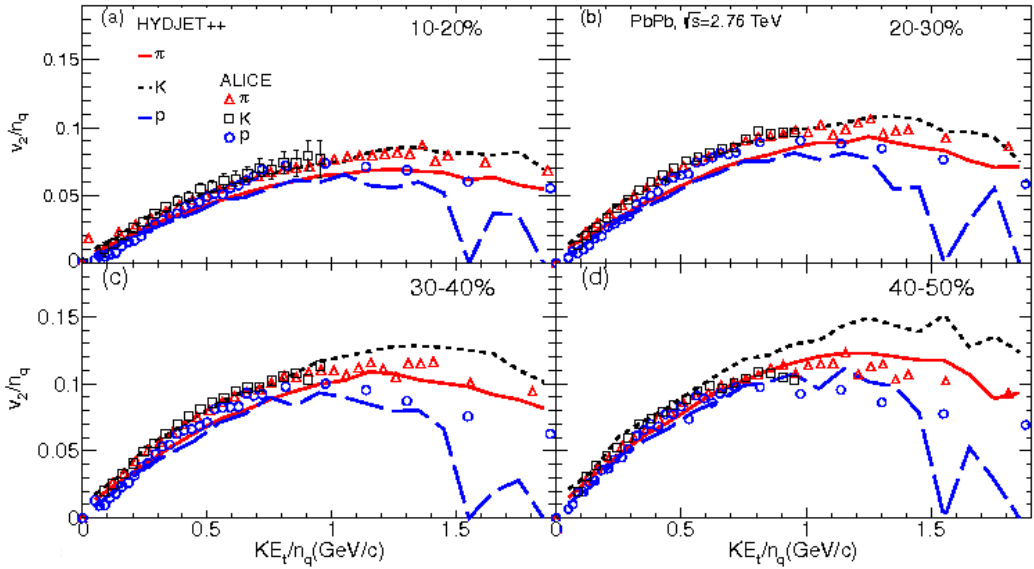}
\caption{\label{fig2}
The $K E_{\rm T}/n_q$ dependence of elliptic flow of pions, kaons and 
protons in Pb+Pb collisions at $\sqrt{s_{\rm NN}} = 2.76$~TeV at 
different centralities. Curves denote the model results, ALICE data
from \cite{alice_ncq} are shown by symbols.}
\enc
\end{figure}

Another important result of the interplay of soft and hard processes is
the worsening of the number-of-constituent-quark (NCQ) scaling for both
elliptic and triangular flow at LHC energies compared to the RHIC ones.
The $v_2(K E_T/n_q) / n_q$ distributions of most abundant hadron 
species, where $K E_T = m_T - m_0$ is the transverse kinetic energy and 
$n_q$ is the number of constituent quarks, are shown in Fig.~\ref{fig2}.        
Experimental results of ALICE collaboration \cite{alice_ncq} are plotted 
onto the calculations as well. We see that the model provides a fair
description of the data. 

The scaling is approximately fulfilled within the 20\% accuracy limit
in the interval $0.1 \leq K E_{\rm T} \leq 0.7$~GeV. The worsening of
the NCQ scaling conditions for $v_2$ at LHC was predicted in HYDJET++ in
\cite{v2_prc09,v2_sqm09}. Our study shows that the effect takes place for
both $v_2$ \cite{v2_prc09,v4_prc13} and $v_3$ \cite{v3_sqm15,v3_arxiv}
distributions. Scaling definitely holds in the hydrodynamic sector of
the model. Moreover, FSI work towards its fulfilment, because many 
light hadrons, especially pions, get the feeddown from the decays of 
heavy resonances. But jets, which are more influential at LHC energies 
compared to that of RHIC, is the main reason causing the NCQ-scaling
violation with increasing collision energy.  

{\it Cross-talk of elliptic and triangular flows.}
The present version of HYDJET++ contains no genuine flow harmonics 
related to the eccentricities of order higher than three. Higher flow
harmonics $v_n,\ n \geq 4$ arise due to nonlinear contributions of 
$v_2$ and $v_3$ \cite{hj_hiharm,v6_prc14}. The detailed comparison
of model predictions with the data shows \cite{hj_hiharm} that HYDJET++
underpredicts the magnitude of $v_4(p_{\rm T})$ - $v_6(p_{\rm T})$ 
signals in central events 0-5\%, but for more peripheral collisions the 
agreement between the model results and the data is much better. It 
means that nonlinear contributions of $v_2$ and $v_3$ to higher 
harmonics dominate over the intrinsic momentum anisotropy $v_n$ caused 
by spatial eccentricity $\varepsilon_n$. Of particular interest is the 
hexagonal flow $v_6$, because in pure hydrodynamic approximation this 
harmonic depends on independent contributions coming from $v_2$ and 
$v_3$ \cite{v6_prc14}:
\beq \ds
v_6 \approx \frac{1}{6} v_2^3 + \frac{1}{2} v_3^2
\label{eq5}
\eeq

\begin{figure}[h]
\begin{minipage}{18pc}
\includegraphics[width=18pc]{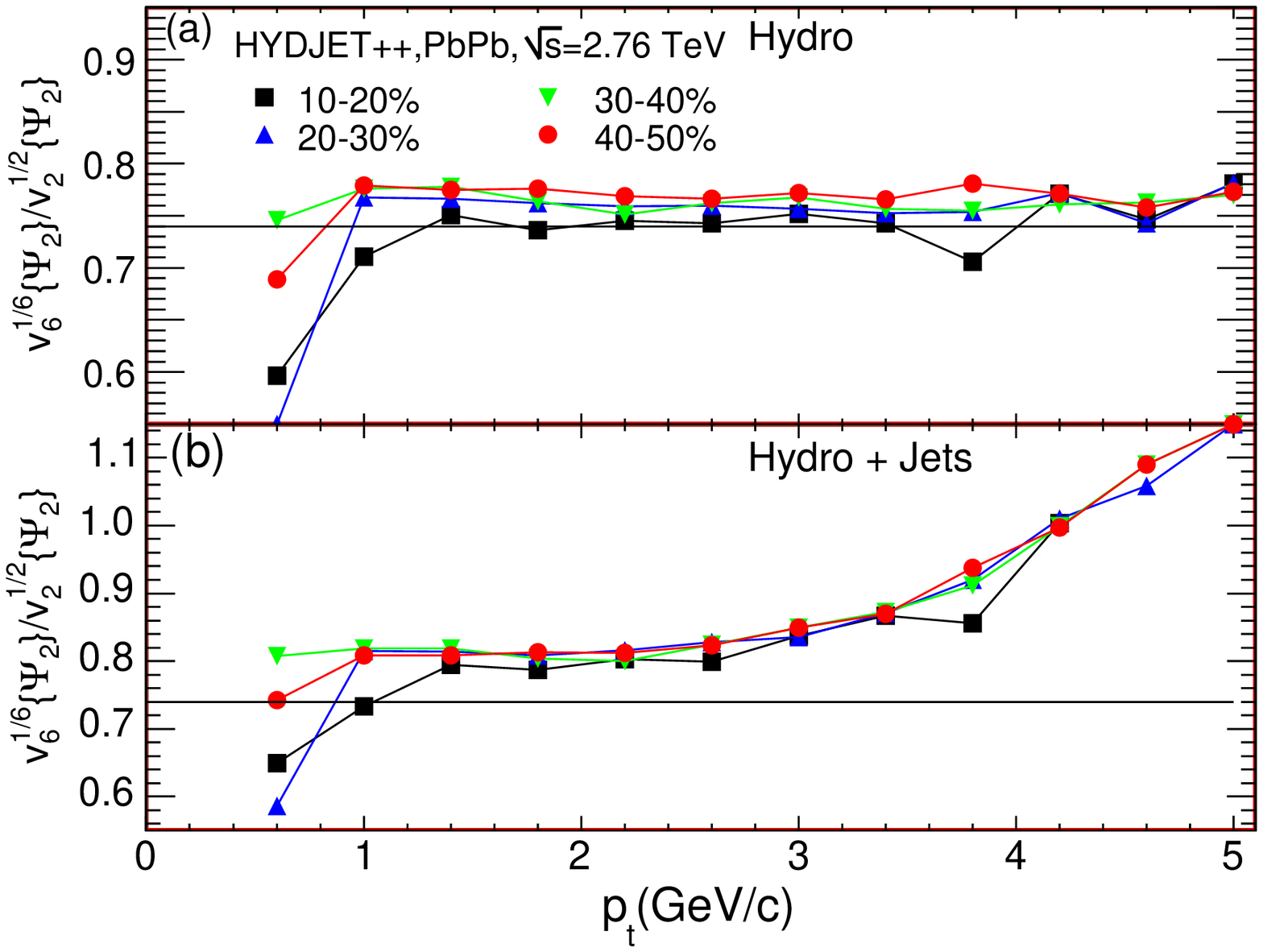}
\caption{\label{fig3}(a) Ratio $v_6^{1/6} / v_2^{1/2}$ in the $\Psi_2$
event plane for charged hadrons from soft processes calculated in 
HYDJET++ for Pb+Pb collisions at 2.76~TeV at several centralities.
(b) The same as (a) but for both soft and hard processes.}
\end{minipage}\hspace{2pc}%
\begin{minipage}{18pc}
\includegraphics[width=18pc]{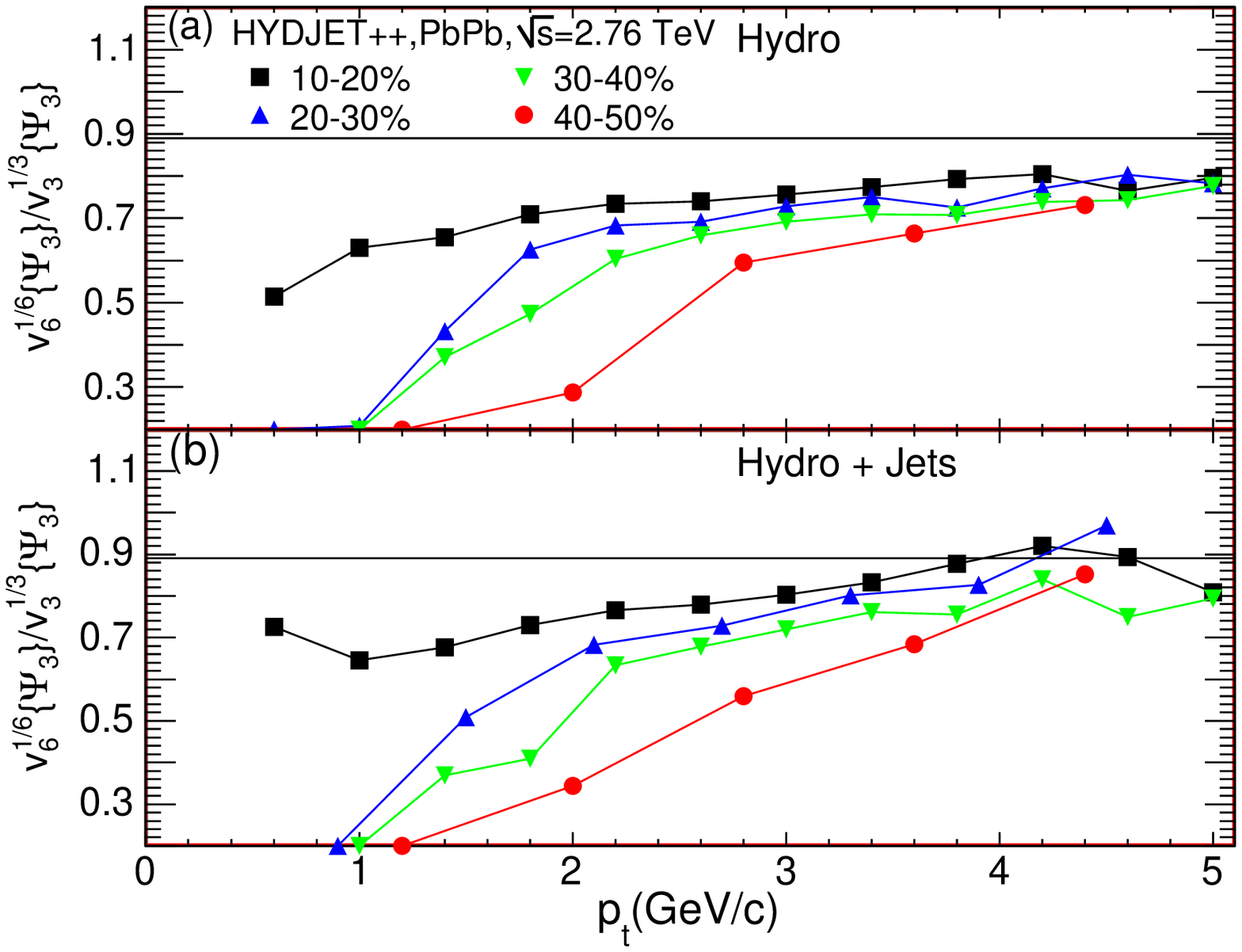}
\caption{\label{fig4}(a) Ratio $v_6^{1/6} / v_3^{1/3}$ in the $\Psi_3$
event plane for charged hadrons from soft processes calculated in
HYDJET++ for Pb+Pb collisions at 2.76~TeV at several centralities.
(b) The same as (a) but for both soft and hard processes.}
\end{minipage}
\end{figure}

The ratio $v_n^{1/n} / v_2^{1/2}$ was proposed in \cite{atlas_scal}
to check the possible scaling trends. We employed this ratio for the
hexagonal flow, and used the $v_6$ defined either in $\Psi_2$ or 
$\Psi_3$ plane, but not in the own $\Psi_6$ plane. Results
\cite{v6_prc14} are displayed in Fig.~\ref{fig3} for
$v_6^{1/6} / v_2^{1/2}$ and in Fig.~\ref{fig4} for $v_6^{1/6} / 
v_3^{1/3}$ distributions.      

For the $v_6^{1/6} / v_2^{1/2}$ distributions in Fig.~\ref{fig3} the 
independence of the ratio on centrality of the collision is observed at
$p_{\rm_T} \geq 1$~GeV/$c$. All curves are on the top of each other.
For the particles produced solely in soft processes this ratio is indeed 
close to $(1/6)^{1/6}$, whereas jets increase it by about 10\% and lead 
to the rise of high-$p_{\rm T}$ tails of the distributions at $p_{\rm T}
\geq 3$~GeV/$c$. For the $v_6^{1/6} / v_3^{1/3}$ ratios the centrality
hierarchy is revealed instead of the scaling. Here the ratio drops as
the reactions become more peripheral. This may be explained by the
significant rise of the elliptic flow with increasing impact parameter
$b$, while the rise of the triangular flow in more peripheral collisions
is not so dramatic. As a result, the event plane $\Psi_6$ becomes more 
correlated with the plane $\Psi_2$ rather than with $\Psi_3$. This is in
line with the experimental observations \cite{atlas_correl}.     

Next interesting issue is the study of two-particle angular correlations.
The two-particle correlation function is typically defined as the ratio 
of pair distribution in the event to the combinatorial background of 
uncorrelated particles. In the flow dominated regime the pair angular 
distribution reads [cf. Eq.(\ref{eq1})]
\beq \ds
\frac{dN^{pairs}}{d\Delta\varphi} \propto 1+ 2\sum_{n=1}^{\infty}
V_n(p_{\rm T}^{\rm tr}, p_{\rm T}^{\rm a})\cos n(\Delta\varphi)~,
\label{eq6}
\eeq
where $\Delta\varphi = \varphi^{\rm tr}-\varphi^{\rm a}$, and indices
``tr" and ``a" indicate the so-called ``trigger" and ``associated"
particle, respectively. The study of angular dihadron correlations in
relativistic heavy-ion collisions revealed the long-range correlations
dubbed ``ridge" \cite{ridge_phenix,ridge_alice}.
Many interesting options have been proposed for the description of the 
ridge phenomenon, for instance Cerenkov gluon radiation or Mach-cone of 
shock waves. The authors of \cite{AlRo_10} suggested that the
triangular flow should be important for understanding of this signal.
HYDJET++ is ideally suited for such a check, because the long-range
correlations in the model appear merely due to the collective flow. 

\begin{figure}[h]
\bec
%\vspace{0.5cm}
\includegraphics[scale=0.45]{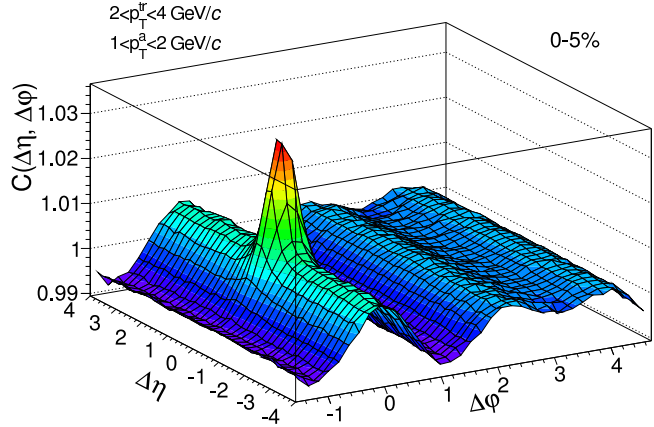}
\caption{\label{fig5}
2D correlation function in HYDJET++ in Pb+Pb collisions
at $\sqrt{s}=2.76$~TeV for $2<p_{\rm T}^{\rm tr}<4$ GeV/$c$ and
$1<p_{\rm T}^{\rm a}<2$ GeV/$c$ for
centrality 0-5\% with both $v_2$ and $v_3$ present.}
\enc
\end{figure}

As was shown in \cite{ridge_prc15}, no long range azimuthal correlations 
at the near-side or away-side $(\Delta\varphi \approx \pi)$ appear in
the case of perfect central collision with $b = 0$, when both elliptic 
and triangular flows are absent. If only the elliptic flow is present,
whereas the triangular flow is switched off, the long range correlations 
start to appear at the both sides. However, only the presence of the 
triangular flow in addition to the elliptic one leads to development of
ridge at near-side and, simultaneously, to formation of characteristic
double-hump structure at the away-side, as seen in Fig.~\ref{fig5}, in 
full agreement with the experimental observations.

{\it Event-by-Event (EbyE) fluctuations.}
The EbyE distributions of harmonics of anisotropic flow in lead-lead
collisions at LHC were studied, e.g., by ATLAS Collaboration in
\cite{atlas_fluct}. The results were obtained after the application of 
the so-called {\it unfolding} procedure \cite{unfold} in order to
extract the ``true" value of the flow vector and get rid of the nonflow
effects caused by the finite event multiplicities, jet fragmentation and
decays of resonances. The procedure is cumbersome, so often people
simply rescale their predictions to make a comparison with the data.
Our analysis shows \cite{fluct_epjc15} that such a simplistic
approach is not always justified.

As an input, one selects the spectra of charged particles with $p_{\rm T}
\geq 0.5$~GeV/$c$ and $|\eta| < 2.5$, corresponding to ATLAS kinematic
cuts. Angular distribution of particles is modified as
\beq \ds
\label{fluct_1}
\frac{dN}{d\varphi} \propto  1 + 2 \sum^{\infty}_{n=1}
V^{\rm obs}_{n} \cos{\left[ n(\varphi - \Psi^{\rm obs}_{n})
\right]} =
1 + 2 \sum^{\infty}_{n=1} \left( V^{\rm obs}_{n,x} \cos{n\varphi} +
V^{\rm obs}_{n,y} \sin{n\varphi} \right)~,
\eeq
with $ V^{\rm obs}_{n}$ being the magnitude of the observed
per-particle flow vector, whereas $ \Psi^{\rm obs}_{n}$ represents
the azimuth of the observed event plane.
Then, the single-particle event-by-event distributions are constructed
\beqar \ds
\label{fluct_2}
\nonumber
& &
V^{\rm obs}_{n} = \sqrt{(V^{\rm obs}_{n,x})^{2} +
(V^{\rm obs}_{n,y})^{2}}~, \\
& &
V^{\rm obs}_{n,x} = V^{\rm obs}_{n} \cos{n\Psi^{\rm obs}_{n}}
= \langle  \cos{n\varphi} \rangle~, \\
\nonumber
& &
V^{\rm obs}_{n,y} = V^{\rm obs}_{n} \sin {n\Psi^{\rm obs}_{n}}
= \langle  \sin{n\varphi} \rangle~.
\eeqar

The two sub-events (2SE) method subdivides the event sample further
into two sub-groups containing charged particles emitted in forward
and backward hemispheres in the c.m. system.
The difference between the EbyE flow vectors (to exclude the collective 
flow) of the two sub-events is
fitted to the Gaussian with the width $\delta_{\rm 2SE} = 2 \delta$,
which enters the response function \cite{atlas_fluct}
\beq \ds
\label{fluct_3}
P ( V^{\rm obs}_{n} \vert V_{n} ) \propto V^{\rm obs}_{n}
\exp \left[ - \frac{( V^{\rm obs}_{n})^{2} + V^{2}_{n}}
{2\delta^{2}} \right]
I_{0}\left( \frac{V^{\rm obs}_{n}V_{n}}{\delta^{2}} \right)~.
\eeq
The obtained response function is then used as an input to the iteration
procedure \cite{unfold} allowing us to find the Bayesian unfolding.
The effects of finite multiplicity and nonflow processes give rise to 
the nonzero value of $\delta_{\rm 2SE}$. The EbyE unfolding procedure
significantly subtracts these contributions and leave the dynamical flow 
fluctuations only.

\begin{figure}[h]
\bec
%\vspace{0.5cm}
\includegraphics[scale=0.80]{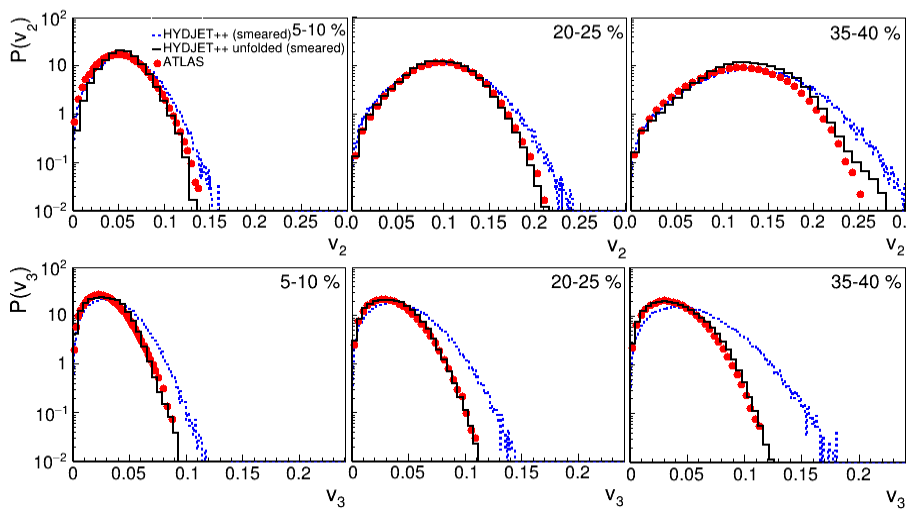}
\caption{\label{fig6}
The probability density distributions of elliptic flow $V_2$
(upper row) and triangular flow $V_3$ (bottom row) in three centrality
intervals: $5-10 \%$ (left), $20-25 \%$ (middle) and $35-40 \%$ (right).
Dashed and solid histograms present the results for simulated HYDJET++
events before and after the unfolding procedure, respectively. The
full circles are the ATLAS data from~\cite{atlas_fluct}.}
\enc
\end{figure}

Figure~\ref{fig6} displays the probability density distributions of
elliptic and triangular EbyE flows obtained in three centrality
intervals: $\sigma/\sigma_{geo} = 5-10\%,\ 20-25\%,\ {\rm and}\ 
35-40\%$. To describe the data we allow for variations of 
$\varepsilon(b)$ and $\varepsilon_3(b)$ \cite{fluct_epjc15}. Now the 
values of both parameters are smeared normally around their previously
fixed values. The width proportionality coefficients are tuned to fit 
the data at a {\it single arbitrary centrality}, say $10-15\%$ or 
$20-25\%$, and the obtained values are used then for all other 
centralities. The initial $P(V_n)$ distributions are
broader than the unfolded ones only in the areas of relatively high
flow values. The agreement of unfolded spectra with the data is very
good. Since the unfolding suppresses strongly the non-flow fluctuations,
Fig.~\ref{fig6} confirms the dynamical origin of the flow fluctuations 
in HYDJET++.

\section{Conclusions}
\label{concl}

Anisotropic flow, azimuthal dihadron correlations and event-by-event
fluctuations of the flow harmonics in Pb+Pb collisions at 
$\sqrt{s_{\rm NN}} = 2.76$~TeV are studied within the hybrid hydro+jets
model HYDJET++. Several features are observed. Hadrons, produced in jet
fragmentation, display very weak flow because of the jet quenching 
effect. These hadrons dominate particle spectrum 
at certain $p_{\rm T}$; the heavier the particle, the larger the 
transverse momentum. Such an interplay of soft and hard processes
explains (i) the breaking of mass ordering of $p_{\rm T}$ distributions
of anisotropic flow; (ii) the falloff of the flow harmonics at 
intermediate transverse momenta; (iii) violation of the NCQ scaling at
LHC energies compared to RHIC ones, because hard processes become more
abundant with rising collision energy. 
The cross-talk of two harmonics, $v_2$ and $v_3$, leads to (i) 
long-range azimuthal dihadron correlations (ridge); (ii) formation of 
the characteristic double-hump structure at the away side; (iii)
nonlinear contributions to higher order harmonics, e.g., to hexagonal
flow, which is correlated with the triangular flow in central collisions
and becomes more correlated with the elliptic flow in more peripheral
ones.
Analysis of EbyE flow fluctuations by means of the unfolding procedure
reveals their dynamical nature in HYDJET++. Its origin is traced to the
correlations between the coordinates and momenta of the hadrons and
the velocities of the fluid cells.  

\ack
LB acknowledges financial support of the Alexander von Humboldt 
Foundation.

\section*{References}
%%%%%%%%%%%%%%%%%%%%%%%%%%%%%%%%%%%%%%%%%%%%%%%%%%%%%%%%%%%%%%%%%%%%%%%%%%%%

\end{document}